\documentclass[aps,prl,twocolumn,groupedaddress,showpacs]{revtex4}

\usepackage{graphicx}

\begin{document}

\title{The holographic principle and the language of genes}

\author{Dirson Jian Li}
\email[]{dirson@mail.xjtu.edu.cn}
\author{Shengli Zhang}

\affiliation{Department of Applied Physics, Xi'an Jiaotong
University, Xi'an 710049, China}

\begin{abstract}
We show that the holographic principle in quantum gravity imposes a
strong constraint on life. The degrees of freedom of an organism can
be estimated according to the theory of Boolean networks, which is
constrained by the entropy bound. Hence we can explain the languages
in protein sequences or in DNA sequences. The overall evolution of
biological complexity can be illustrated. And some general
properties of protein length distributions can be explained by a
linguistic mechanism.
\end{abstract}


\maketitle
\date{}

\section{Introduction}

\noindent The general principles in non-living systems play
significant roles in living systems. How do the principles in
gravity theory or in quantum mechanism impact on our understanding
of life? An organism can not keep active without the supply of
energy due to the first law in thermodynamics. And it can not live
long without the supply of minus entropy due to the second law in
thermodynamics. But it seems that there are no direct effects of the
relativity principle or the uncertainty principle on life. We found
that the holographic principle, which is likely only one of several
independent conceptual advances needed for progress in quantum
gravity
\cite{HolographicPrinciple}\cite{HolographicPrinciple_2}\cite{HolographicPrinciple_3}\cite{HolographicPrinciple_4}\cite{HolographicPrinciple_5},
profoundly constraints the forms of life and substantially impacts
on the evolution of life.

The holographic principle states that there is a precise, general
and surprisingly strong limit on the information content of
spacetime regions. The number of quantum states in a spatial region
is bounded from above by the surface of the region measured in the
unit of four-fold Planck areas. This entropy bound is a strong
constraint on any theory about our universe. If this principle is
true, field theory or string theory, where there are infinite
degrees of freedom, can not be the ultimate theory. And if this
principle can be applied to the phenomenon of life, the degrees of
freedom in a living system will also be constrained. From this point
of view, the principles in relativity or in quantum theory constrain
life in an alternative way. The holographic principle indicates that
there is a strict relationship between the information storage
capacity of the space and the complexity of any organism wherein.
Such a basic idea can be illustrated by a simple example. Whatever a
living system with $n$ degrees of freedom is, we can conclude that
it can never exist in a universe with a horizon area less than $4 n
l_p^2$, where $l_p$ is the Planck length.

In this paper, we estimated the immense degrees of freedom for
living systems according to the theory of gene regulatory networks
and Boolean networks \cite{GRN} \cite{Kauffman}. We found a
contradiction between the possible degrees of freedom of living
systems and the maximum information storage capacity in the observed
universe. Then we reconciled this contradiction in terms of the
causality between the possible sequences of macromolecules for the
actual living systems, which is equivalent to the existence of
language of genes. We propose evidences of language of genes and we
can explain the outline of protein length distributions by a
linguistic mechanism of generation of protein sequences. We can also
explain the leaps in the evolution of biological complexity
according to the entropy bound.

\section{Immense degrees of freedom in living systems}

Information properly bridges biology and physics \cite{information
biophy}\cite{information biophy_2}\cite{information biophy_3}, which
gives deep insights into the nature of life. With the development of
genetics, we know that the gene regulatory networks play significant
roles in development and evolution of life \cite{GRN}. Based on the
theory of self-organization, Kauffman proposed a general theory of
Boolean networks to describe the gene regulatory networks, where the
interactions between genes can represented by Boolean operations
between the nodes of the network \cite{Kauffman}. Thus, the degrees
of freedom of a living system can be estimated by the number of
states of the corresponding Boolean network. Proteins are the
elementary units in the activities of life. So a living organism can
be represented by a dynamical system of all the proteins in its
body. We denote the set $\mathcal {P}$ as all possible protein
sequences with a cutoff of protein length $l$. Proteins are chains
concatenated by $20$ amino acids. So there are $m=\Sigma_{k=1}^l
20^k$ elements in the set $\mathcal {P}$. We define a Boolean
network $\mathcal {N}$ as the Boolean network whose nodes are
elements of $\mathcal {P}$ (Fig. 1a). According to the definition of
Boolean networks, there are two states for each node of a Boolean
network: ``on'' or ``off'' \cite{Kauffman}. A state of $\mathcal
{N}$ represent that some nodes are ``on'' while the others are
``off''. So a proteome can be represented by a state of $\mathcal
{N}$, where only the nodes corresponding to protein sequences in the
proteome are ``on''. The state space $\mathcal {S}$ consists of all
possible states of $\mathcal {N}$ whose number is
\begin{equation}n_s' = 2^m.\end{equation} An actual species can be
represented by a point in $\mathcal {S}$. The evolution of a species
can be illustrated by a trajectory in $\mathcal {S}$ (Fig. 2b). As a
preliminary consideration, the degrees of freedom of a living system
can be estimated by the logarithm of number of states
\begin{equation}d' \sim \ln n_s' \sim 20^l \ln2,\end{equation} which
we will reconsider later on.

According to the holographic principle, we can calculate that the
information in the observed universe is about \cite{Susskind book}
\begin{equation}I_{univ}=10^{122}\ bits. \end{equation}
This value is too large for non-living systems. For example, the
information of black body cosmic background photons is about
$10^{90}$ bits, which may be the largest degrees of freedom for
possible non-living systems. But it is still much less than
$I_{univ}$. The remaining information storage capacity in our
universe has not been wasted however for there being living systems.
The degrees of freedom for living systems are so immense that may
exceed the maximum information storage capacity in the observed
universe.

The structure of chains of genetic macromolecules essentially
provides immense degrees of freedom for living systems, because the
number of possible protein sequences can be as large as $20^l$. For
a living system, the degrees of freedom may be equivalent to that of
the observed universe if the protein length is about $n^*=94$ amino
acids. Interestingly, the most frequent protein length for the life
on our planet is about $n^*$. The immense degrees of freedom of
living systems originate from the great number of possible sequences
in $\mathcal {N}$. Most of the degrees of freedom come from the
states of $\mathcal {N}$ in which about half the nodes are ``on''.
On the other hand, the degrees of freedom can also come from the
states in which only a minority of nodes are ``on''. Our living
systems belong to the latter case, where there are only thousands of
proteins in actual proteomes.

\section{Entropy bound and the causality of sequences}

\begin{figure}
\centering{
\includegraphics[width=50mm]{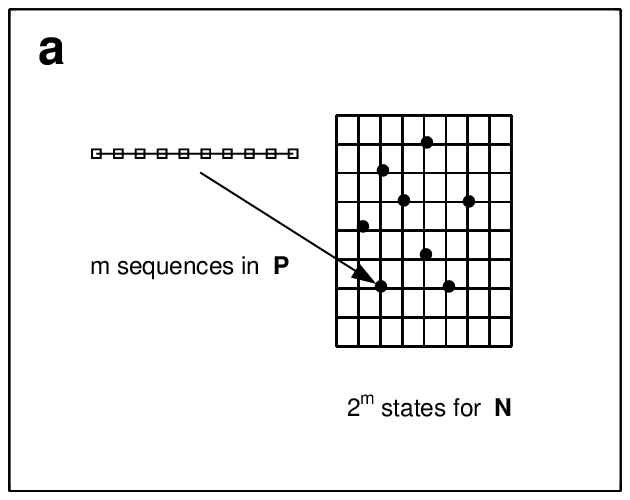}
\includegraphics[width=50mm]{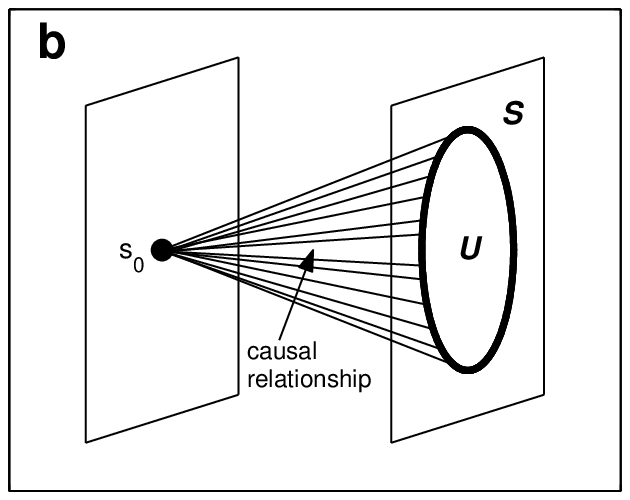}
} \label{fig1} \caption{\small {\bf Boolean networks and the causal
relationship between the macromolecular sequences.} {\bf a,} The
nodes of the Boolean network ${\mathcal N}$ consist of all possible
protein sequences in ${\mathcal P}$ with length less than $l$ amino
acids. For each node, there are two states ``on" or ``off", so there
are $2^m$ states for ${\mathcal N}$. The state $s_0$ is a state of
$\mathcal {N}$ in which some nodes are ``on'' (represented by black
dots). {\bf b,} Only a part of the states in ${\mathcal N}$ may have
biological meaning in an actual living system. The number of states
in ${\mathcal N}$ may exceed $I_{univ}$, but the number of states in
${\mathcal U}$ can not be greater than $I_{univ}$.}
\end{figure}

The estimate of immense degrees of freedom of a living system in the
above, however, seriously contradicts the holographic principle if
we consider the actual life around us. The average protein length in
a proteome ranges about from $250$ amino acids to $550$ amino acids,
and a certain number of proteins are longer than thousands of amino
acids. According to the preliminary estimate, the degrees of
freedoms for the actual living systems on our planet will be much
larger than the maximum degrees of freedoms in the observed universe
$I_{univ}$. We have to reconcile the contradiction between the
preliminary estimate of degrees of freedom of living systems and the
conclusion of the holographic principle. If the holographic
principle is not invalid, we must find ways to shrink the
preliminary estimate of the degrees of freedom of living systems.

We introduce the causality between the states of $\mathcal {N}$ to
reveal the additional constraint on the degrees of freedom by the
entropy bound. At the beginning of the evolution on the planet, the
first living system may be denoted as an inertial state $s_0$. When
the degrees of freedom of $\mathcal {N}$ is greater than $I_{univ}$,
not all the states of $\mathcal {N}$ can have causal relationship
with $s_0$ unless the holographic principle is untrue. We define the
set $\mathcal {U}$ as all the states that have causal relationship
with $s_0$, which has $n_s$ states and is only a proper subset of
$\mathcal {S}$ (Fig. 1b). The nodes of $\mathcal {U}$ constitute
$\mathcal {L}$, which is a subset of $\mathcal {P}$. An actual
living system at present corresponds to a dynamic system evolving
only in the state space $\mathcal {U}$ and a meaningful protein
sequence in biology must belong to $\mathcal {L}$. The degrees of
freedom of a living system, therefore, can be defined by the number
of states in $\mathcal U$:
\begin{equation}d \equiv \ln n_s, \end{equation} where $n_s$ is much
less than $n_s'$ and $d$ can be rightly less than $I_{univ}$.

\section{The language required by the holographic
principle}

\begin{figure}
\centering{
\includegraphics[width=40mm]{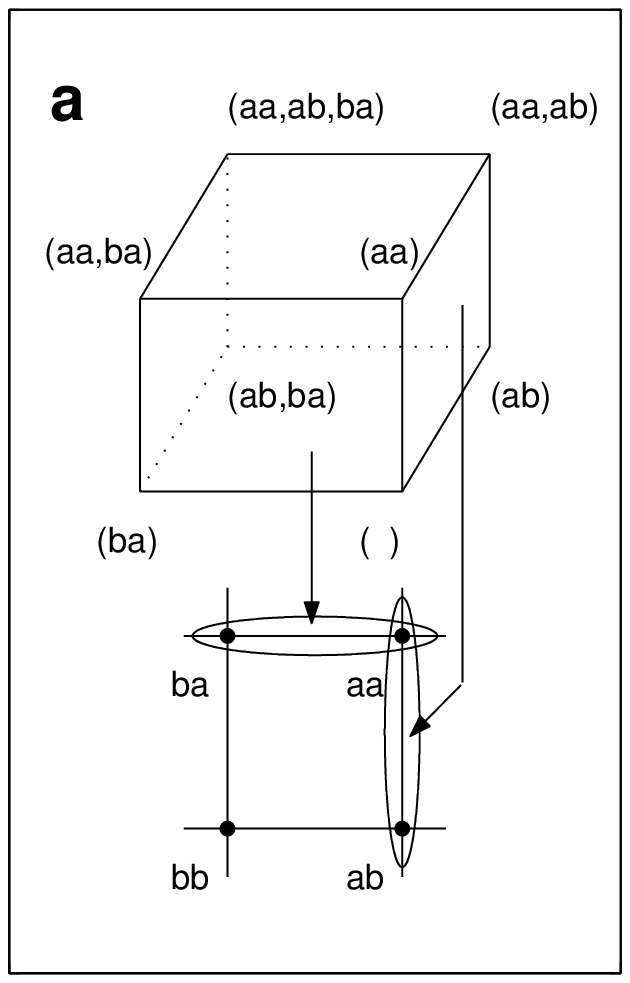}
\includegraphics[width=40mm]{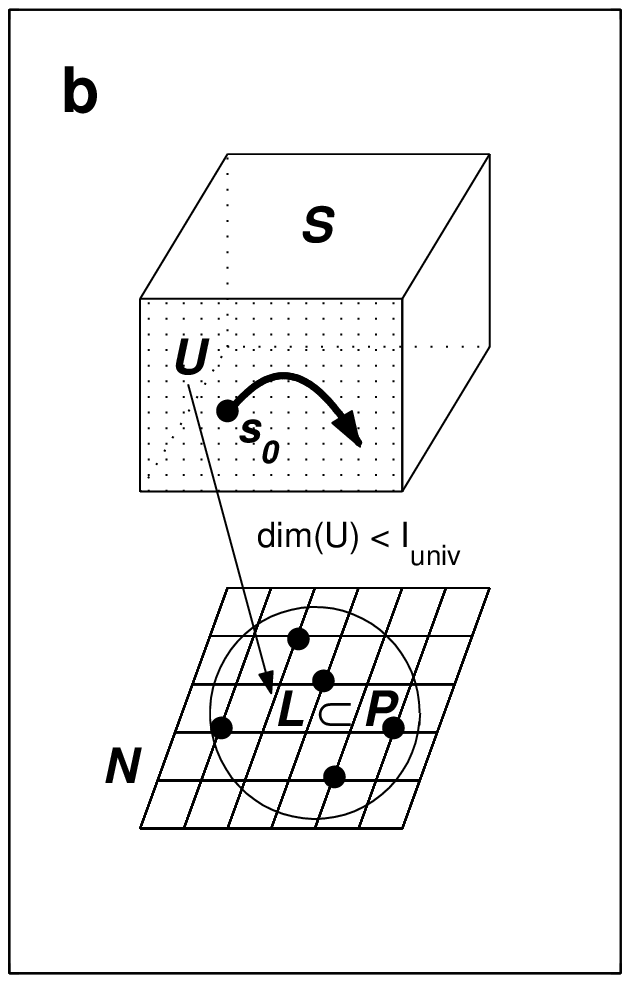}
} \label{fig1} \caption{\small {\bf The finite degrees of freedom
require language of sequences.} {\bf a,} A toy model can explain the
necessity of the language in sequences. Suppose that the entropy
bound requires that the information can not be greater than $2$ bits
in a tiny universe. There is no living system corresponding to the
Boolean network with $3$ nodes $aa$, $ab$ and $ba$. We can choose a
subset $aa$ and $ab$ as the nodes of an available Boolean network,
which corresponds to an actual living system in this universe. Thus
we obtain a language consisting of $2$ words: $aa$ and $ab$. {\bf
b,} Only the states in the set $\mathcal U \subset \mathcal S$ has
causal relationship with the inertial state $s_0$ due to the entropy
bound. We can obtain a language $\mathcal L$ of the sequences, which
is a subset (dots on $\mathcal N$) of $\mathcal P$. The number of
elements in $\mathcal L$ must be less than $I_{univ}$.}
\end{figure}

The causality provides a physical explanation to distinguish a part
of sequences $\mathcal {L}$ from all possible sequences $\mathcal
{P}$. Not all the amino acid chains or base chains are meaningful in
biology. According to the theory of formal language, a language is
defined by a subset of all the sequences concatenated by letters in
a given alphabet \cite{Formal language}. The choice of a subset
$\mathcal {L}$ from $\mathcal {P}$ is a natural way to define a
formal language (Fig. 2). The protein or DNA languages originate in
the constraint on the degrees of freedom of life by the entropy
bound. The alphabet of protein language consists of $20$ amino
acids, and the alphabet of the language of genes consists of $4$
bases. The arrangement of the letters in the sequences should be
determined by some grammars. Although there are various entropy
bounds, there is no difference for the requirement of finite degrees
of freedom in life and the requirement of the language of genes for
all the theories. To some extent, the language of genes is a
consequence of the principles in quantum gravity. The phenomenon of
life is constrained strictly by the entropy bound. The requirement
of the order of sequences by the grammars can not be explained in
the context of classical physics because the degrees of freedom of
life can be infinite.

The ability of speaking for human beings is determined by genes.
That we can communicate with each other instinctively can be
attributed to our common genes. The human language can be viewed as
a transformation of cell language \cite{natural language DNA
language}. The information storage capacity of a natural language
can also be estimated by the similar calculation in the above. For
instance, we estimated that there are up to $I_{human}=26^{l'}$ bits
of information can be written in a language with $26$ letters and
the length of words in the language is $l' \sim 10$, which is much
less than the protein length. In this sense, the natural language is
simpler than the language of genes. The value $I_{human}$ is much
less than the information in the observed universe $I_{univ}$. So
the description of the universe by natural language is always a
simplified version of the actually complex world. Interestingly,
there were not rare cases to reach the same goal by different routes
in the history of natural sciences, such as, Riemannian Geometry and
general relativity, or the theory of bundles and gauge theories.
Such encounters may come from that all the descriptions in different
subjects have a common ultimate theory of all the information in the
universe, although we can not understand all the details of the
world by only one subject.

\section{The language of genes and underlying order in sequences}

\begin{figure}
\centering{
\includegraphics[width=80mm]{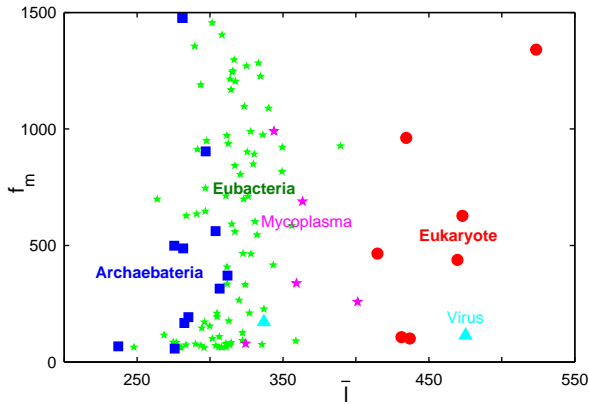}
} \label{fig3} \caption{\small {\bf Relationship between the average
protein length $\bar{l}$ and the frequency $f_m$ of the highest peak
of discrete fourier transformation of protein length distribution.}
The distribution of the species from three domains likes a rainbow.
Even for the group of closely related species such as mycoplasmas
(belonging to eubacteria), their distribution also form an ``arch''
of the rainbow. This is a strong evidence for the underlying
mechanism of the protein length distributions.}
\end{figure}

Several attempts have been made over the past three decades to
combine linguistic theory with biology \cite{ProteinLinguistice}
\cite{Searls}. The distribution of the number of occurrences of
protein domains in a genome can be a good fit of the power-law
distribution known as Zipf's law in linguistics, and we can
distinguish between the protein linguistics and the language of
genes according to the theory of formal language \cite{Searls}. So
the experimental observations support the existence of languages in
the sequences of macromolecules. On one hand, they are required by
the holographic principle. On the other hand, they are consequences
of the evolution of life at the molecular level
\cite{EvolCode}\cite{EvolCode_2}\cite{EvolCode_3}\cite{AA
Chronology}. The alphabets of amino acids or bases formed at the
beginning of life. And genetic code developed and fixed in the early
stage of evolution. All these factors can determine whether a
sequence is permitted in a life, which is equivalent to the role of
grammars at the molecular level.

We found a strong evidence of the underlying mechanism in the
organization of amino acids in protein sequences by studying the
correlations between protein length distributions, which indicates
the languages in the protein sequences. The protein length
distribution corresponds to a vector \begin{equation}{\mathbf
D}=(D(1), D(2), ..., D(g),...D(c)),\end{equation} where there are
$D(g)$ proteins with length $g$ in the complete proteome of a
species and $c=3000$ is the cutoff of protein length. Our data of
the protein length distributions are obtained from the data of $106$
complete proteomes in the database Predictions for Entire Proteomes
\cite{PEP}. The discrete fourier transformation of the protein
length distribution is: \begin{equation} \tilde{D}(f) =
\frac{1}{\sqrt{c}} \sum _{g=1} ^c D(g) e^{2 \pi i (g-1)(f-1)/c}
\end{equation} Let $f_m$ denotes the frequency of the highest peak
$\tilde{D}(f_m)$ in the discrete fourier transformation of the
protein length distribution for a species. We found that there is an
interesting relationship between the frequency $f_m$ and the average
protein length $\bar{l}$ of species. The distribution of species in
$\bar{l}-f_m$ plane shows an regular pattern: the species in the
three domains (Archaebacteria, Eubacteria and Eukaryotes) gathered
in three rainbow-like arches respectively (Fig 3). This pattern
strongly indicates the intrinsic correlation among the protein
length distributions, which can never achieve if the protein length
distributions are stochastic. The periodic-like fluctuations in the
protein length distribution \cite{Periodic distribution} may also
originate in the underlying mechanism of generation of protein
sequences.

\section{Explanation of the order in protein sequences}

We propose a model to reveal the underlying mechanism in the protein
sequences according to tree adjoining grammar \cite{TAG}. In the
model, protein sequences can be generated by tree adjoining
operations, i.e., substituting the initial tree or auxiliary trees
into to each other by identifying the inner nodes (Fig. 4a)
\cite{TAG}. There is only one variant $t$ in the model, which is the
probability of substitutions in the adjoining operations and denotes
different species. A certain number of proteins can be generated
when $t$ is fixed, hence we obtain a protein length distribution by
the model (Fig. 4b). The properties of protein length distributions
can be explained by the simulation. The outline and the fluctuations
of the simulated protein length distribution agree with the actual
protein length distributions in principle.

We show that there is a close relationship between the protein
length distributions and grammar rules. The fluctuations in the
distributions are determined by the grammar rules. The same grammar
rule corresponds to the same distribution. If changing grammar
rules, we obtain different outlines and fluctuations of
distribution. This result suggests that the fluctuations in actual
protein length distributions are intrinsic properties of certain
species and may infer the underlying mechanism on the order of
protein sequences.

\begin{figure}
\centering{
\includegraphics[width=80mm]{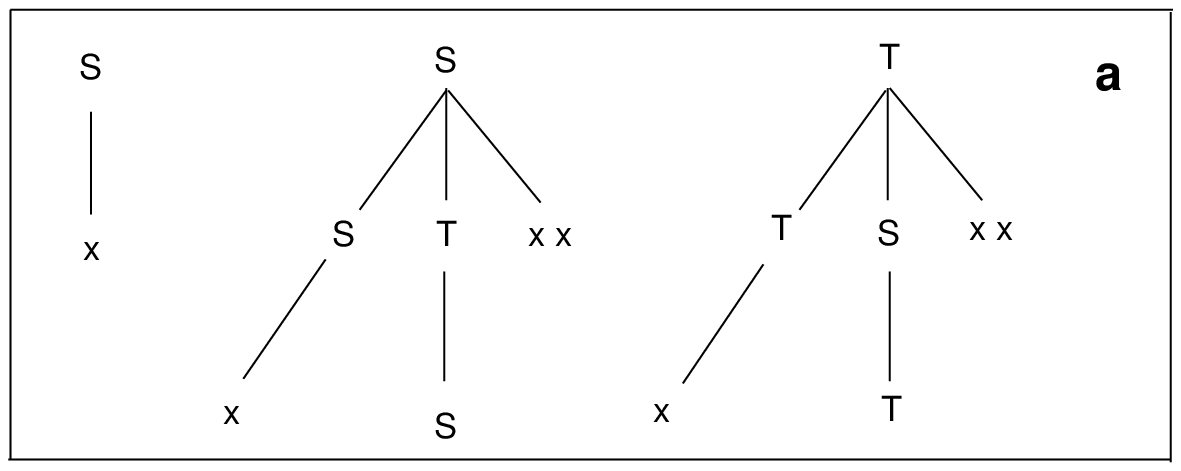}
\includegraphics[width=80mm]{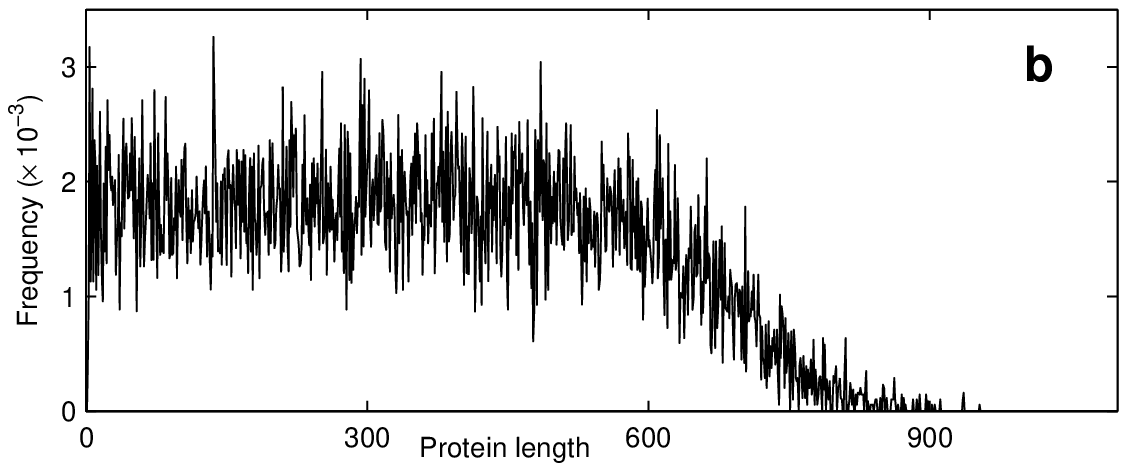}}
\label{fig1} \caption{\small {\bf Simulation of protein length
distributions by a linguistic model.} {\bf a,} The tree adjoining
grammar. There are one initial tree and two auxiliary trees, where
$S$ and $T$ are inner nodes and $x$ or $x\ x$ are leaves which
represent the amino acids. {\bf b,} The simulation of protein length
distribution by the tree adjoining grammar. The properties of
protein length distributions such as the outline and fluctuations
can be simulated by the linguistic model.}
\end{figure}

\section{The macroevolution of biological complexity}

The evolution of complexity of life is not a linear course of
increment \cite{evolution of complexity}\cite{evolution of
complexity_2}. The entropy bound can also explain the leaps in the
evolution of biological complexity. Consequently we can outline the
macroevolution of life. The gene regulatory networks are
accelerating networks
\cite{AcceleratingNetworks}\cite{AcceleratingNetworks Arxiv}.
According to this theory, the evolution of complexity of any
accelerating networks has to be slowed down and will stop at an
upper limit of complexity. Hence there must be upper limits of
complexity in both of the evolution of biological complexity for
prokaryotes and eukaryotes, where the entropy bound is a natural
upper limit. The whole evolution of biological complexity can be,
therefore, divided into three steps: the evolution of unicellular
life, the evolution of multicellular life and the evolution of
society of human beings. The Cambrian explosion divided the first
two steps. And we found that the evolution of multicellular life has
reached its upper limit because the maximum non-coding DNA content
is near to $1$ at present. The civilization of human beings
appeared, which can be taken as an alternative form of biological
complexity. The entire evolution of biological complexity should be
governed by a universal mechanism of evolution. The universal
language of genes in species may harmonize the evolution of life in
the biosphere.
\begin{acknowledgments}

We thank Hefeng Wang, Liu Zhao, Yachao Liu and Lei Zhang for
valuable discussions. Supported by NSF of China Grant No. of
10374075.

\end{acknowledgments}

\end{document}